\documentclass[aps,longbibliography,showpacs,twocolumn,superscriptaddress]{revtex4-1}
\usepackage{amsmath,amssymb,amsfonts,bm}
\usepackage{graphicx}
\usepackage{epstopdf}
\usepackage{dcolumn}
\usepackage{mathrsfs}
\usepackage{bbold}
\usepackage{dsfont}
\usepackage{float}
\usepackage[colorlinks=true,linkcolor=blue,citecolor=blue, urlcolor=blue,bookmarks=false]{hyperref}

\usepackage{tgtermes}
\usepackage{multirow}

\bibpunct{[}{]}{,}{n}{}{}

\begin{document}

\title{Magnon corner states in twisted bilayer honeycomb magnets}
\date{\today }
\author{Chun-Bo Hua}
\altaffiliation{These authors contributed equally to this work.}
\affiliation{School of Electronic and Information Engineering, Hubei University of Science and Technology, Xianning 437100, China}
\affiliation{Laboratory of Optoelectronic Information and Intelligent Control, Hubei University of Science and Technology, Xianning 437100, China}
\author{Feiping Xiao}
\altaffiliation{These authors contributed equally to this work.}
\affiliation{School of Physics and Electronics, Hunan University, Changsha 410082, China}
\author{Zheng-Rong Liu}
\altaffiliation{These authors contributed equally to this work.}
\affiliation{Department of Physics, Hubei University, Wuhan 430062, China}
\author{Jin-Hua Sun}
\affiliation{Department of Physics, Ningbo University, Ningbo 315211, China}
\author{Jin-Hua Gao}
\affiliation{School of Physics and Wuhan National High Magnetic Field Center, Huazhong University of Science and Technology, Wuhan 430074, China}
\author{Chui-Zhen Chen}
\affiliation{Institute for Advanced Study and School of Physical Science and Technology, Soochow University, Suzhou 215006, China}
\author{Qingjun Tong}\email{tongqj@hnu.edu.cn}
\affiliation{School of Physics and Electronics, Hunan University, Changsha 410082, China}
\author{Bin Zhou}\email{binzhou@hubu.edu.cn}
\affiliation{Department of Physics, Hubei University, Wuhan 430062, China}
\author{Dong-Hui Xu}
\email{donghuixu@cqu.edu.cn}
\affiliation{Department of Physics and Chongqing Key Laboratory for Strongly Coupled Physics, Chongqing University, Chongqing 400044, People's Republic of China}
\affiliation{Center of Quantum Materials and Devices, Chongqing University, Chongqing 400044, People's Republic of China}

\begin{abstract}
	
The study of symmetry-protected topological phases of matter has been extended from fermionic electron systems to various bosonic systems. Bosonic topological magnon phases in magnetic materials have received much attention because of their exotic uncharged topologically protected boundary modes and the potential for dissipationless magnonics and spintronic applications. Here, we establish twisted bilayer honeycomb magnets as a platform for hosting second-order topological magnon insulators (SOTMIs) without fine-tuning. We employ a simple, minimal Heisenberg spin model to describe misaligned bilayer sheets of honeycomb ferromagnetic magnets with a large commensurate twist angle. We found that the higher-order topology in this bilayer system shows a significant dependence on the interlayer exchange coupling. The SOTMI, featuring topologically protected magnon corner states that go beyond the conventional bulk-boundary correspondence, appears for ferromagnetic interlayer couplings, while the twisted bilayer exhibits a nodal phase in the case of antiferromagnetic interlayer coupling. At last, relevance to twisted bilayer CrI$_3$ is also discussed.
\end{abstract}

\maketitle

\emph{\color{red}Introduction.}---After the discovery of time-reversal invariant topological insulators, symmetry-protected topological phases of matter have been an exciting and cutting-edge area of research in condensed matter physics \cite{Hasan2010RMP,Qi2011RMP,Bansil2016RMP,Haldane2017RMP,Wen2017RMP,Wolfle2018RPP}. Remarkably, symmetry-protected topological phases are not unique to electronic systems, and have been identified in various bosonic systems either, where topological magnon phases have received special attention for their potential applications in spintronics \cite{kondo2020nonhermiticity,li2020topological,doi:10.1063/5.0041781,mcclarty2021topological}. Up to now, fruitful topological magnon phases including insulating and semimetallic phases, have been investigated both theoretically and experimentally  \cite{Onose297,PhysRevLett.104.066403,PhysRevB.87.144101,PhysRevB.89.134409,PhysRevB.90.024412,PhysRevLett.115.106603,PhysRevLett.117.187203,
PhysRevB.94.174444,PhysRevB.97.134411,2020arXiv201009945X,PhysRevLett.115.147201,PhysRevB.101.100405,PhysRevB.95.014422,
PhysRevLett.117.227201,Owerre_2016JAP,Owerre_2016JPCM,Owerre_2017JPC,pantaleon2018effects,zhang2020interplay,PhysRevB.97.081106,PhysRevX.8.041028,
Cao_2015JPCM,PhysRevB.102.214421,PhysRevB.95.014435,PhysRevApplied.9.024029,PhysRevLett.125.217202,PhysRevB.87.174402,PhysRevB.87.174427,PhysRevB.90.104417,
PhysRevB.87.024402,PhysRevB.97.174413,PhysRevB.96.224414,PhysRevB.97.180401,PhysRevB.97.140401,PhysRevB.100.144401,PhysRevB.99.041110,
PhysRevB.94.075401,PhysRevLett.117.157204,PhysRevB.95.224403,kondo2021dirac,PhysRevB.103.014407,PhysRevLett.122.187203,
Li_2016NC,PhysRevLett.119.247202,PhysRevB.95.014418,PhysRevX.8.011010,yao2018topological,lu2021topological,zhu2021topological}. Recently, the concept of higher-order topological insulators~\cite{Benalcazar2017Science,Schindler2018SA,Langbehn2017PRL,Song2017PRL,PhysRevLett.124.036803} has been extended to magnonic systems as well \cite{Sil_2020,PhysRevLett.125.207204,PhysRevB.104.024406,Li_2019,PhysRevB.101.184404,PhysRevB.104.L060401}. The hallmark feature of an $n$th-order magnon topological insulator in $d$ dimensions is the existence of protected gapless magnon states at its $(d-n)$-dimensional boundaries, which go beyond the celebrated bulk-boundary correspondence. For example, a second-order topological magnon insulator (SOTMI) with magnon corner states is realized in a ferromagnetic (FM) Heisenberg model on a two-dimensional (2D) breathing kagome lattice \cite{Sil_2020}, a magnonic quadrupole topological insulator hosting magnon corner states can appear in 2D antiskyrmion crystals \cite{PhysRevLett.125.207204}, and an SOTMI with 1D chiral hinge magnons is predicted to be realized in 3D stacked honeycomb magnets \cite{PhysRevB.104.024406}. All these existing magnonic higher-order topological insulators require significant Dzyaloshinskii-Moriya interaction, whereas the Dzyaloshinskii-Moriya interaction is a typically weak effect in most magnetic materials~\cite{dzyaloshinsky1958thermodynamic,Moriya1960}.

\begin{figure*}[!t]
	\centering
	\includegraphics[width=16cm]{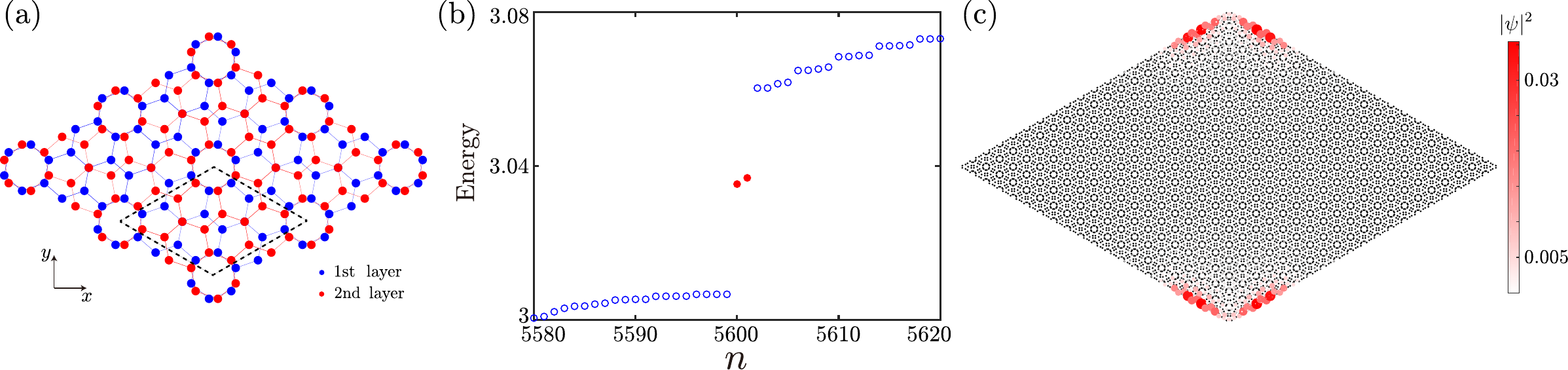} \caption{(a) Schematic illustration of the TBHM lattice with the large commensurate angle $\theta=21.78^{\circ}$. Blue dots represent the first layer lattice, and red dots represent the second layer lattice. The blue (red) lines denote the first (second) layer intralayer nearest-neighbor bonds. The rhombus shaped area composed of the black thick dashed lines represents a single moir\'e unit cell. (b) Magnon energy spectrum of the Hamiltonian~(\ref{H}) of the rhombus-shaped TBHM system versus the eigenvalue index $n$. Red dots mark the in-gap magnon corner states. (c) The spatial distribution of the probability density of the two in-gap states in (b). The color map shows the values of the probability density. We take the exchange coupling ratio $J_{\bot }/J=0.2$, and the lattice site number $N=11200$.}%
\label{fig1}
\end{figure*}

In recent years, 2D twisted van der Waals materials have emerges as a versatile platform for studying exotic and elusive states of matter, following the discovery of unconventional superconductivity \cite{Cao_2018nature1} and the Mott insulator \cite{Cao_2018nature2} in twisted bilayer graphene (TBG) with magic angles~\cite{Bistritzer12233}. TBG has been shown to yield a series of fascinating correlated and topological phenomena \cite{Bistritzer12233,Lu_2019nature,Yankowitz_2019science,Sharpe_2019science,PhysRevLett.122.106405,Stepanov_2020nature,Serlin_2019science,
Xie_2019nature,Jiang_2019nature,Zondiner_2020nature,Wong_2020nature}. Besides the intrinsic fragile topology \cite{PhysRevX.8.031089,PhysRevLett.121.126402,PhysRevLett.123.036401,PhysRevX.9.021013} of the nearly flat bands, higher-order band topology has been subsequently identified in TBG as well \cite{PhysRevLett.123.216803,PhysRevLett.126.066401,PARK2021260,chew2021higherorder}. Meanwhile, researchers have turned their attention to twisted bilayer honeycomb magnets (TBHMs) analogous to twisted bilayer graphene, and revealed rich magnetic phases caused by moir\'e patterns as well as intriguing moir\'e magnetic excitations in TBHMs \cite{Hejazi10721,Ghader_2020SR,PhysRevB.102.094404,Zhu_2021}. Moreover, Moir\'e magnetism has been reported in twisted bilayer $\rm CrI_3$ \cite{doi:10.1126/science.abj7478,Xu_2021NN,xie2022twist} in a very recent experiment. Inspired by the recent theoretical and experimental developments in twisted 2D magnets, it is tempting to ask that whether higher-order topological magnon insulators can occur in TBHMs.

In this work, we reveal that an SOTMI can be realized in the TBHM at the large commensurate angle, without requiring the Dzyaloshinskii-Moriya interaction. We adopt a simple, minimal spin model which consists of two honeycomb FM layers with the nearest-neighbor intralayer exchange interaction coupled by the FM or antiferromagnetic (AFM) interlayer exchange coupling, to describe the TBHMs with collinear order. For our purpose, we assume that the interlayer exchange coupling is sufficiently weak compared to the intralayer Heisenberg interaction. Therefore, the out-of-plane collinear magnetic order is favored under weak interlayer coupling. Furthermore, we obtain an effective magnon Hamiltonian in terms of the Holstein-Primakoff transformation to bosonize the spin model. Based on numerical diagonalization, we show that the FM interlayer coupling can give rise to an energy gap associated with the nontrivial higher-order topology characterized by a mirror winding number, resulting in an SOTMI in the TBHM. The SOTMI supports two in-gap magnon corner states localized at mirror symmetric corners. The magnon corner states are robust against symmetry-preserving disorder. In contrast, in the case of AFM interlayer coupling, the TBHM system remains gapless and has magnon Dirac dispersion. Our work, together with these works on higher-order topology in twisted photonic \cite{PhysRevB.103.214311,yi2021strongly} and acoustic \cite{wu2021observation} materials, suggest a natural strategy to realize bosonic higher-order topological insulators.

\emph{\color{red}Model.}---We consider a twisted AA-stacked bilayer honeycomb magnets with the commensurate angle $\theta=21.78^{\circ}$, whose spins are localized at the hexagon vertices marked by red and blue dots, as shown in Fig.~\ref{fig1}(a). The spin Hamiltonian is formulated on the twisted bilayer honeycomb lattice, which reads
\begin{equation}
H=-J\sum_{\left\langle i,j\right\rangle ,l}\mathbf{S}_{i,l}\cdot \mathbf{S}%
_{j,l}-J_{\bot }\sum_{\left\langle i,j\right\rangle }\mathbf{S}_{i,2}\cdot
\mathbf{S}_{j,1},
\end{equation}
where the first and second terms represent the nearest-neighbor intralayer and nearest-neighbor interlayer Heisenberg interactions, respectively. $\mathbf{S}_{i,l}=(S^{x}_{i,l},S^{y}_{i,l},S^{z}_{i,l})$ is the spin vector operator at site $i$ on layer $l=1,2$, and the summation runs over nearest-neighbor sites $\left\langle i,j\right\rangle$. $J>0$ denotes the FM intralayer interaction, and $J_{\bot }$ is a tunable parameter in TBHMs, which is positive for the FM interlayer coupling while negative for the AFM coupling. Here, $J_{\bot }$ only couples the sites of the first layer with the sites of the second layer that are positioned directly next to them. In the Supplemental Material (SM)~\cite{SM}, we also show the results when including the spatially modulated remote interlayer couplings.

Noticing that the twisted bilayer system is constructed by twisting the bilayer magnets with respect to the collinear axis at the hexagonal center, where the lower layer rotates counterclockwise $\theta/2$ and the upper layer rotates clockwise $\theta/2$, respectively. The system belongs to $D_6$ point group~\cite{PhysRevLett.123.216803} with concurrent spatial and spin rotations. More specifically, the system is invariant under the action of either of the symmetry operators: $C_{6z}$, sixfold rotation about the out-of-plane $z$ axis, and $C_{2x/2y}$, twofold rotation about the in-plane $x/y$ axis.

In the FM case, the classical ground state is represented by the uniform state $\mathbf{S}_{i,l}\equiv S \mathbf{\hat{z}}$, where the spins point along the $+z$ direction. In the ordered phase supported at sufficiently low temperatures, we obtain an effective magnon Hamiltonian through the linear spin-wave theory. Using the Holstein-Primakoff transformation \cite{PhysRev.58.1098}
\begin{align}
S_{i}^{+}=&S^{x}_{i}+iS^{y}_{i}\simeq \sqrt{2S}d_{i}, \nonumber \\
 S_{i}^{-}=&S^{x}_{i}-iS^{y}_{i}\simeq \sqrt{2S} d_{i}^{\dag }, \\
 S_{i}^{z}=&S-d_{i}^{\dag }d_{i}, \nonumber
\end{align}
and neglecting magnon-magnon interactions, the spin Hamiltonian can be transformed into a quadratic magnon Hamiltonian
\begin{align}
\label{H}
H=&3JS \sum_{i,l}d_{i,l}^{\dag }d_{i,l}-JS\sum_{\left\langle i,j\right\rangle ,l}( d_{i,l}^{\dag}d_{j,l}+\text{H.c}.)  \\
&+J_{\bot}S \sum_{\left\langle i,j\right\rangle }[(d_{i,2}^{\dag }d_{i,2}+d_{j,1}^{\dag }d_{j,1})-(d_{i,2}^{\dag }d_{j,1}+\text{H.c.})], \nonumber
\end{align}
where $d_{i}^{\dag }$ ($d_{i}$) is the bosonic creation (annihilation) operator.
In subsequent calculations, the energy unit is set as the intralayer Heisenberg interaction amplitude $J$. In addition, the lattice constant of monolayer and the interlayer spacing between layers are both set to $1$.

\begin{figure*}[t]
\centering
	\includegraphics[width=17cm]{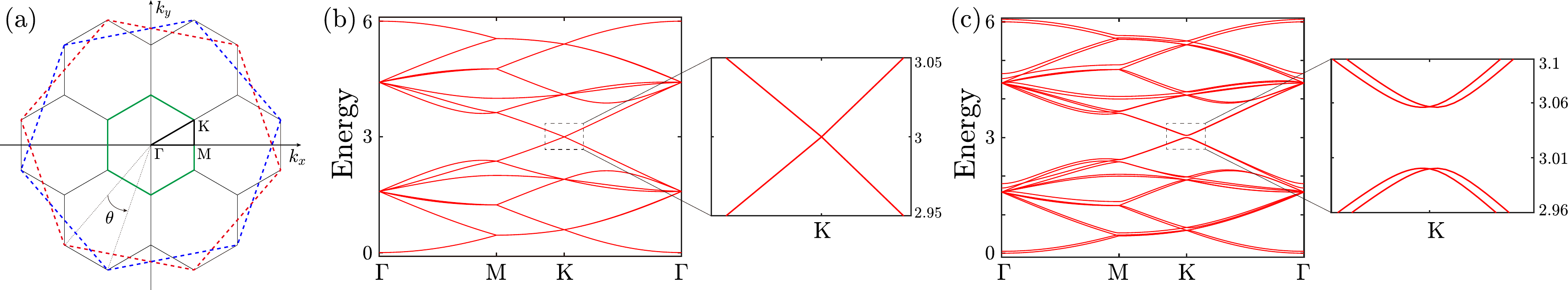} \caption{(a) Moir{\'e} Brillouin zone. The dashed red (first layer) and blue (second layer) large hexagons show the unfold Brillouin zones of individual layers, respectively, and the blue hexagon corresponds to red hexagon for the twist angle $\theta=21.78^{\circ}$. The first moir{\'e} Brillouin zone of the bilayer is shown by the central (green) thick solid hexagon. The next several Brillouin zones of the TNHM are depicted by the six surrounding (black) thin solid hexagons. The magnon band structures are calculated along the high symmetry path specified by the triangle $\Gamma\text{M}\text{K}$ (black). The symmetry points are $\Gamma=(0,0)$, $\text{M}=(2\sqrt{7}\pi/21,0)$, and $\text{K}=(2\sqrt{7}\pi/21,2\sqrt{21}\pi/63)$. (b) Magnon band structure of the TBHM with $\theta=21.78^{\circ}$ for the FM interlayer interaction amplitude $J_{\bot }/J=0$. (c) Magnon band structures for the FM interlayer interaction amplitude $J_{\bot }/J=0.2$. The zoom-ins demonstrate the dispersion around the $K$ point.}%
\label{fig2}
\end{figure*}

\emph{\color{red}Magnon corner states induced by the FM interlayer coupling.}---To diagnose the higher-order topology of the TBHM system with the FM interlayer coupling, we first calculate the magnon energy spectrum of the TBHM with a rhombus boundary preserving the twofold rotational symmetry. By numerically diagonalizing the magnon Hamiltonian Eq.~(\ref{H}) in real space, we plot the magnon energy spectrum in Fig.~\ref{fig1}. It is found that the magnon energy spectrum shows an energy gap, and even more interestingly two in-gap states reside in the energy gap~[shown in Fig.~\ref{fig1}(b)]. As shown in Fig.~\ref{fig1}(c), the two in-gap states are symmetrically localized at the top and bottom corners of the rhombus, respectively, which are two mirror symmetric corners. Note that, due to the finite size effect, the two magnon corner states are not degenerate, and a small gap due to the hybridization of two corner states exists. The twofold symmetric in-gap corner states are a hallmark feature of the SOTMI in the TBHM, which are associate with a mirror winding number as demonstrated later in this paper.

In the SM~\cite{SM}, we demonstrate the two mirror symmetric magnon corner states still exist when including remote intralayer and interlayer FM exchange couplings. Moreover, we also show that six magnon corner states appear when considering a finite hexagon-shaped TBHM sample in the SM. These magnon corner states are six-fold rotation symmetric and localized at the six corners of the sample.

\emph{\color{red}Magnon band structures and mirror winding number.}---To gain more insight into the origin of magnon corner states, we study the bulk band structure of the TBHM and its topology. At the commensurate angle $\theta=21.78^{\circ}$, the original translational symmetry of AA-stacked bilayer honeycomb is broken, but the moir{\'e} translational symmetry can be defined to display the periodicity of superlattices. Thereby, under the Fourier transformation, we obtain a $28 \times 28$ magnon Hamiltonian in the $k$ space, which reads $H=\sum_{\mathbf{k}} \Psi _{\mathbf{k}}^{\dag } H_{\mathbf{k}} \Psi _{\mathbf{k}}$ with the basis $\Psi _{\mathbf{k}}^{\dag }=( c_{\mathbf{k},1}^{\dag },\cdots ,c_{\mathbf{k},28}^{\dag })$. The concrete expression of $H_{\mathbf{k}}$ and more details are given in Ref.~\cite{SM}.

The magnon band structures of the TBHM system along a high symmetry line of the moir{\'e} Brillouin zone [see Fig. \ref{fig2}(a)] obtained by numerically diagonalizing $H_{\mathbf{k}}$ are shown in Fig.~\ref{fig2}.
For comparison, we also depict the magnon band structure of the TBHMs system in the absence of interlayer interaction $J_{\bot }/J=0$ in Fig.~\ref{fig2}(b). It is found that the magnon band structure shows linearly dispersive bands around the moir{\'e} Brillouin Zone corner $\mathbf{K}$, which is identical to the monolayer honeycomb magnet. Subsequently, we present the magnon band structures of the TBHM system with a finite FM interlayer exchange interaction in Fig.~\ref{fig2}(c), where we set $J_{\bot }/J=0.2$ echoing the magnon energy spectrum of the system with open boundary conditions shown in Fig.~\ref{fig1}. We conclude that the finite FM interlayer coupling opens a sizable energy gap at the point $\mathbf{K}$. The energy gap is topologically non-trivial since magnon corner states emerge within it when the open boundary condition is imposed.

To further illustrate the topological properties of the magnon band structures, we utilize the $\mathbb{Z}_2$ mirror winding number \cite{PhysRevLett.123.216803} as a topological invariant to characterize the higher-order magnon topology. The mirror winding number $\nu$ is defined in a mirror-invariant line $\Gamma$-$M$-$\Gamma$ in the moir{\'e} Brillouin zone, where $C_{2x}$ symmetry is preserved. Then we can decompose the Hamiltonian $H_{\mathbf{k}}(  k_{x},0)$, which situates at this mirror-invariant line $\Gamma$-$M$-$\Gamma$, into two decoupled parts $H_{\pm}(k_{x}) $ by projecting $H_{\mathbf{k}}( k_{x},0) $ onto the subspace formed by the eigenvectors that correspond to the mirror eigenvalues $\pm1$ of the mirror operator $C_{2x}$. The mirror winding number $\nu$ is defined as $\nu=\nu_{+}=\nu_{-}$ $(\operatorname{mod}2)$, where $\nu_{\pm}$ is the winding number in the two subsectors.

The mirror winding number can be calculated by the Wilson loop method \cite{Benalcazar2017Science,Benalcazar2017PRB,PhysRevLett.123.216803}. The Wilson loop operator $W_{\pm}$ is considered in the mirror-invariant line $\Gamma$-$M$-$\Gamma$, where $k_{i}(k_{f})$ is the initial (final) point of the loop. We define the element of a matrix $F_{x,k_{i}}^{\pm}$ as $[  F_{x,k_{i}}^{\pm}]  ^{mn}=\langle u_{\pm,k_{i}+\Delta k}^{n}|u_{\pm,k_{i}}^{n}\rangle$, where $\Delta k$ is the spacing of momentum in the loop and $\vert u_{\pm,k_{x}}^{n}\rangle $, for $n=1...N_{occ}$, are the occupied Bloch functions of a crystal with $N_{occ}$ occupied energy bands. Next, the Wilson loop operator can be expressed as $W_{\pm}=F_{k_{f}-\Delta k}^{\pm}F_{k_{f}-2\Delta k}^{\pm}\cdots F_{k_{i}+\Delta k}^{\pm}F_{k_{i}}^{\pm}$. Therefore, the winding number reads
\begin{equation}
\nu_{\pm}=\frac{1}{i\pi}\log( \det[  W_{\pm}] ).
\end{equation}
In this work, the magnon corner states appear when the mirror winding number is $\nu=\nu_{\pm}=+1$, which confirms the topological origin of the magnon corner states localized at the mirror invariant corners of the TBHM.

\begin{figure}[pb]
	\includegraphics[width=8.4cm]{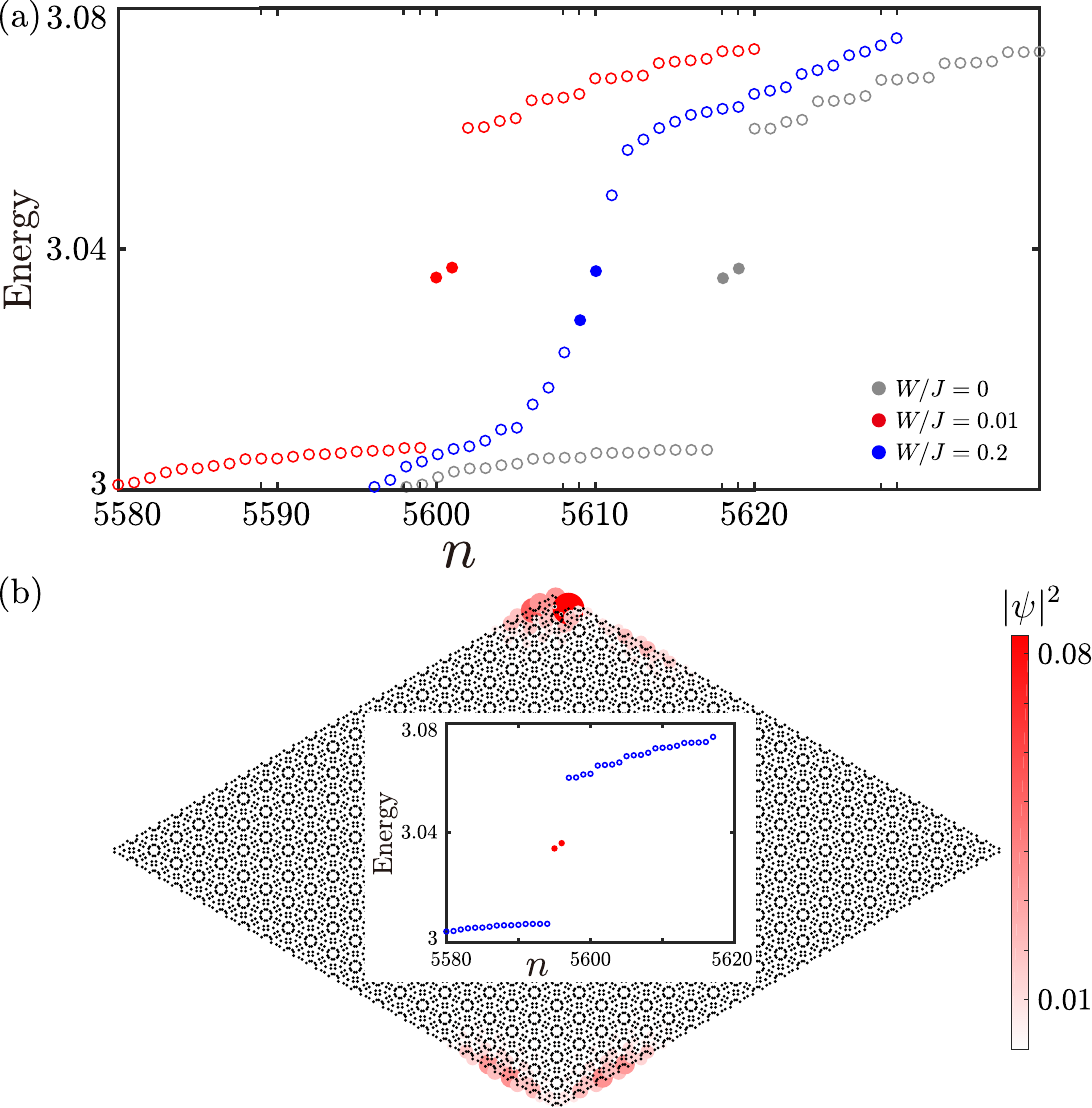} \caption{(a) Magnon energy spectrum of the total Hamiltonian $H+H_{z}$ versus the eigenvalue index $n$. Red dots mark all the in-gap states. For comparison, we also plot the magnon energy spectrum (shown in grey circles and dots) without disorder ($W/J=0$) and the magnon energy spectrum (shown in blue circles and dots) with strong disorder ($W/J=0.2$). We take the model parameters $J_{\bot }/J=0.2$, $W/J=0.01$ and lattice site number $N=11200$. (b) Magnon energy spectrum of the Hamiltonian $H$ on the TBHMs system with a defect versus the eigenvalue index $n$, and the probability density of the two in-gap states, where $J_{\bot }/J=0.2$ and the lattice site number $N=11190$. Red dots mark all the in-gap states. The color map shows the values of the probability density.}%
\label{fig3}
\end{figure}

\emph{\color{red}Stability of corner states.}---Here, we use a random magnetic field to examine the robustness of the magnon corner states. The Zeeman term induced by the random magnetic field along $z$-direction can be expressed as
\begin{equation}
H_{z}=-\sum_{i,l}B_{i,l}\mathbf{S}_{i,l}\cdot \mathbf{z}.
\end{equation}
The random magnetic field is $B_{i}=W \omega_{i}$, where $\omega_{i}$ is the uniform random variable chosen from $[-0.5,0.5]$ and $W$ is the disorder strength. Within the framework of the linear spin-wave theory and via Holstein-Primakoff transformation, the Zeeman field term can be transformed into $H_{z}=\sum_{i,l}B_{i,l}d_{i,l}^{\dag} d_{i,l}$, where all elements situate at the diagonal of the magnon Hamiltonian matrix, resembling the on-site chemical potential disorder known from the electronic version.

In Fig.~\ref{fig3}(a), we demonstrate the magnon energy spectrum versus the eigenvalue index $n$ for different disorder strength. We find that the two in-gap magnon corner states remain stable in the case of weak disorder, where we take the disorder strength as $W=0.01$. While the in-gap magnon corner states are destroyed and pushed into bulk states by the strong disorder with the disorder strength ($W=0.2$) shown in Fig.~\ref{fig3}(a). In addition, we also reveal the robustness of the magnon corner states by introducing a local defect into the rhombus boundary sample at the top corner, where the defect is constructed by removing $10$ sites. In the presence of defects, we plot the magnon energy spectrum and the spatial probability density of the two in-gap states in Fig.~\ref{fig3}(b). We can see that the two in-gap states are still stable and localized around the original two corners, although the spatial distribution becomes mirror asymmetric. Meanwhile, in the SM~\cite{SM}, we also show that the six-fold rotation symmetric magnon corner states are robust against random magnetic fields and local defects.

\emph{\color{red}AFM interlayer exchange coupling.}---We briefly discuss the case of AFM interlayer Heisenberg interaction in this section. We assume that the spins of the first (second) layer are polarized along the positive (negative) $z$-direction. Similarly, using the Holstein-Primakoff transformation and neglecting magnon-magnon interactions, an effective magnon Hamiltonian $H_{\rm AFM}$ is obtained.
\begin{figure}[t]
\centering
	\includegraphics[width=8.5cm]{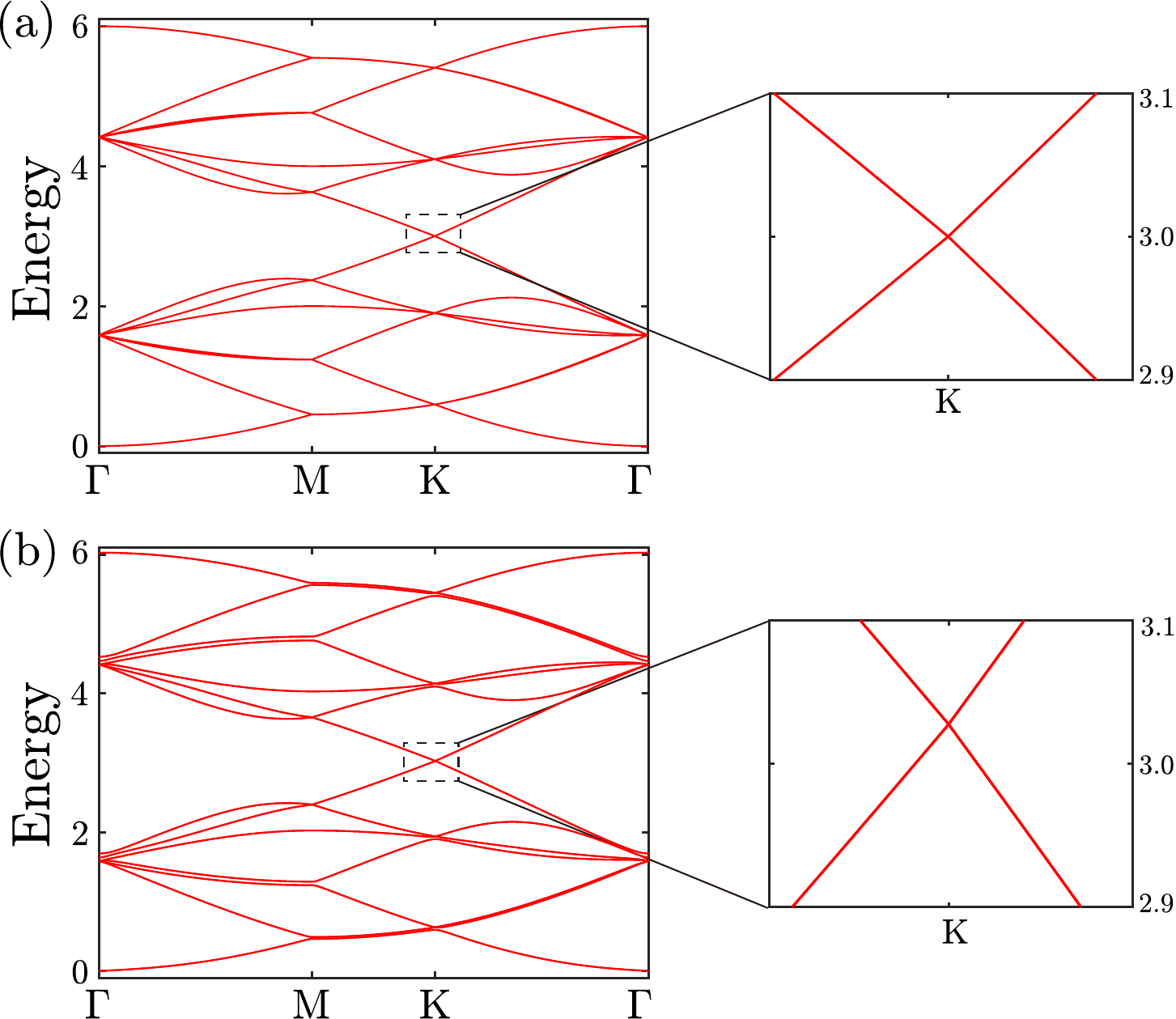} \caption{(a) Magnon band structures of TBHMs with $\theta=21.78^{\circ}$ for the AFM interlayer interaction amplitude $J_{\bot }/J=0$, and (b) Magnon band structures for the AFM interlayer interaction amplitude $J_{\bot }/J=-0.2$. The zoom of the region near K point indicated by a dashed black box in the left panel of (a) and (b) are shown in the right panel.}%
\label{fig4}
\end{figure}
Again by using the Fourier transformation, we obtain a $28 \times 28$ magnon Hamiltonian in the $k$ space, which can be expressed as $H_{\rm AFM}=\sum_{\mathbf{k}} \psi _{\mathbf{k}}^{\dag } H_{\rm AFM}(\mathbf{k}) \psi _{\mathbf{k}}$ with the basis $\psi_{\mathbf{k}}=(c_{\mathbf{k},1},...,c_{\mathbf{k},14},c_{\mathbf{-k},15}^{\dag},...,c_{\mathbf{-k},28}^{\dag})^{T}$. To obtain the AFM magnon band structures, we use a paraunitary Bogoliubov transformation $\psi_{\mathbf{k}}=R(\mathbf{k})\phi_{\mathbf{k}}$ to diagonalize the $k$-space magnon Hamiltonian as $R(\mathbf{k})^{\dag}H_{\rm AFM}(\mathbf{k})R(\mathbf{k})\!=\!D$, where $D$ is a diagonal matrix and $\phi_{\mathbf{k}}=(1,...,1,-1,...,-1)^{T}$. More details of the Hamiltonian in the AFM case are given in the Ref.~\cite{SM}.

We display the AFM magnon band structures of the TBHM system in Fig.~\ref{fig4}. Figure~\ref{fig4}(a) shows the magnon band structure of the TBHMs system in the absence of the interlayer interaction, where we take the parameter $J_{\bot }/J=0$. The magnon band structures of the TBHM system with a finite AFM interlayer interaction is shown in Fig.~\ref{fig4}(b), where we set $J_{\bot }/J=-0.2$. It is found that, in contrast to the FM case, the AFM magnon energy bands of the system keep gapless, and the linear dispersion is stable, regardless of whether AFM interlayer interactions exist.

At last, it is necessary to point out that the ferromagnetic interlayer coupling induces the intervalley scattering of Dirac magnons thus opens the energy gap, which is similar to the fermionic systems~\cite{PhysRevB.81.161405,PhysRevLett.123.216803}. On the contrary, the antiferromagnetic coupling only involves the Dirac magnons within the same valley, which leaves them gapless.

\emph{\color{red}Conclusion and Discussion.}---In this work, we have investigated higher-order topology of in the TBHM with the large commensurate angle $\theta=21.78^{\circ}$. Based on a simple, minimal spin model, we found the FM interlayer coupling hybrids the Dirac bands of two individual honeycomb layers and thus opens a topological band gap characterized by the mirror winding number, leading to an SOTMI in the TBHM. The SOTMI supports hallmark magnon corner states, which are robust against weak random magnetic fields and local defects. In contrast, in the case of AFM interlayer coupling, the linearly dispersive magnon bands remain gapless.

Our theory is immediately testable considering rapid experimental progress on 2D twisted magnets \cite{doi:10.1126/science.abj7478,Xu_2021NN,xie2022twist}. Monolayer Chromium triiodide (CrI$_3$) is a 2D ferromagnet with honeycomb lattice~\cite{huang2017layer}. For a $\theta=21.78^{\circ}$ twisted bilayer CrI$_3$, our first-principles calculations show that interlayer FM interation is always favored, although its magnitude depends on the interlayer stacking configurations [See the calculations in the SM~\cite{SM}, also, references~\cite{Kresse1996,PBE1994,Perdew1996,Klimes2011,Liechtenstein1995,xu2018interplay,PhysRevB.101.060404,PhysRevLett.127.166402} therein]. Accordingly, we predict that twisted bilayer CrI$_3$ is a promising setup to realize the SOTMI featuring magnon corner states.

\emph{\color{red}Acknowledgments.}---D.-H.X. was supported by the NSFC (under Grant Nos. 12074108 and 12147102) and the Natural Science Foundation of Chongqing (Grant No. CSTB2022NSCQ-MSX0568). B.Z. was supported by the NSFC (under Grant No. 12074107), the program of outstanding young and middle-aged scientific and technological innovation team of colleges and universities in Hubei Province (under Grant No. T2020001) and the innovation group project of the natural science foundation of Hubei Province of China (under Grant No. 2022CFA012). C.-Z.C. was funded by the NSFC (under Grant No. 11974256) and the NSF of Jiangsu Province (under Grant No. BK20190813). C.-B.H was supported by the Doctoral Research Start-Up Fund of  Hubei University of Science and Technology (under Grant No. BK202316).

\bibliographystyle{apsrev4-1-etal-title_6authors}
\bibliography{bibfile}

\pagebreak
\widetext
\clearpage

\setcounter{equation}{0}
\setcounter{figure}{0}
\setcounter{table}{0}
\makeatletter
\renewcommand{\figurename}{FIG.}
\renewcommand{\theequation}{S\arabic{equation}}
\renewcommand{\thetable}{S\arabic{table}}
\renewcommand{\thefigure}{S\arabic{figure}}

\begin{center}
\textbf{\large Supplemental Material to: ``Magnon corner states in twisted bilayer honeycomb magnets''}
\end{center}



In this supplemental material, we first give a generic spin model of the twisted bilayer honeycomb magnets (TBHMs) with remote ferromagnetic (FM) exchange couplings and the magnon corner states in Sec.~I. Then, in Sec.~II we present the magnon Hamiltonian in the $k$ space and the magnon band structures. Next, we briefly discuss the case of antiferromagnetic (AFM) interlayer Heisenberg interaction in Sec.~III. We show the magnon corner states in a sample with regular hexagonal boundary shape in Sec.~IV. At last, we give the first principle calculations for twisted bilayer CrI$_3$.

\section{FM Long-range spin model and magnon corner states}
In this section, we consider a generic spin Hamiltonian of the TBHMs and study the magnon corner states. The long-range FM spin Hamiltonian reads
\begin{equation}
H=-J_{p}\sum_{i\neq j,l}\mathbf{S}_{i,l}\cdot \mathbf{S}_{j,l}-J_{\perp
}^{q}\sum_{i\neq j}\mathbf{S}_{i,2}\cdot \mathbf{S}_{j,1},
\label{H1}
\end{equation}
with
\begin{eqnarray}
J_{p} &=&J_{0}\exp \left( -\frac{a_{p}-a_{0}}{\delta }\right),  \\
J_{\perp }^{q} &=&J_{\perp }^{0}\exp \left( -\frac{b_{q}-b_{0}}{\delta }\right).
\end{eqnarray}
The first term of the Hamiltonian represents the intralayer Heisenberg interaction described by the FM exchange coupling $J_{p}$, and the second term denotes the interlayer Heisenberg interaction described by the FM exchange coupling $J_{\perp}^{q}$. $J_{0}$ is the nearest-neighbor (NN) intralayer FM exchange coupling. $J_{\perp}^{0}$ is the NN interlayer FM exchange coupling. $a_{p}$ is a distance that controls the intralayer coupling strength, where $a_{1}=1$ for the NN distance, $a_{2}=\sqrt{3}$ for the next nearest-neighbor (NNN) distance, and $a_{3}=2$ for the next next nearest-neighbor (NNNN) distance. $b_{q}$ is a distance that controls the interlayer coupling strength, where $b_{1}=1$ for the NN interlayer distance, $b_{2}$ for the NNN interlayer distance, and $b_{3}$ for the NNNN interlayer distance. $a_{0}=1$ is the distance of two NN sites within a single layer, $b_{0}=1$ is the interlayer spacing between layers, and $\delta$ is the decay length of the coupling term. In the main text, we only keep the NN intralayer and interlayer interactions in the Hamiltonian, where $J_{p}=J_{0}=J$ and $J_{\perp}^{q}=J_{\perp}^{0}=J_{\perp}$.

\begin{figure}[htbp]
	\centering
	\includegraphics[width=12cm]{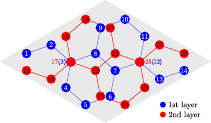} \caption{Schematic illustration of a unit cell for the TBHMs lattice with the commensurate angle $\theta=21.78^{\circ}$. The first (second) layer lattice sites are marked by the filled blue (red) circles. A unit cell contains $28$ lattice sites, and each site is marked with a number.}%
	\label{figS1}
\end{figure}

Schematic illustration of a unit cell for the TBHMs lattice with the commensurate angle $\theta=21.78^{\circ}$ is shown in Fig.~\ref{figS1}. The NN interlayer coupling connects the two sites are placed directly to each other, such as the site $3$ of the first layer and the site $17$ of the second layer in Fig.~\ref{figS1}. The NNN interlayer coupling connects the slightly displaced sites relative to the stacked sites, such as the site $9$ of the first layer and the site $20$ of the second layer in Fig.~\ref{figS1}. As shown in Fig.~\ref{figS1}, the site $8$ in the first layer and the site $22$ in the second layer are coupled by the NNNN interlayer coupling.

\begin{figure}[htbp]
	\centering
	\includegraphics[width=12cm]{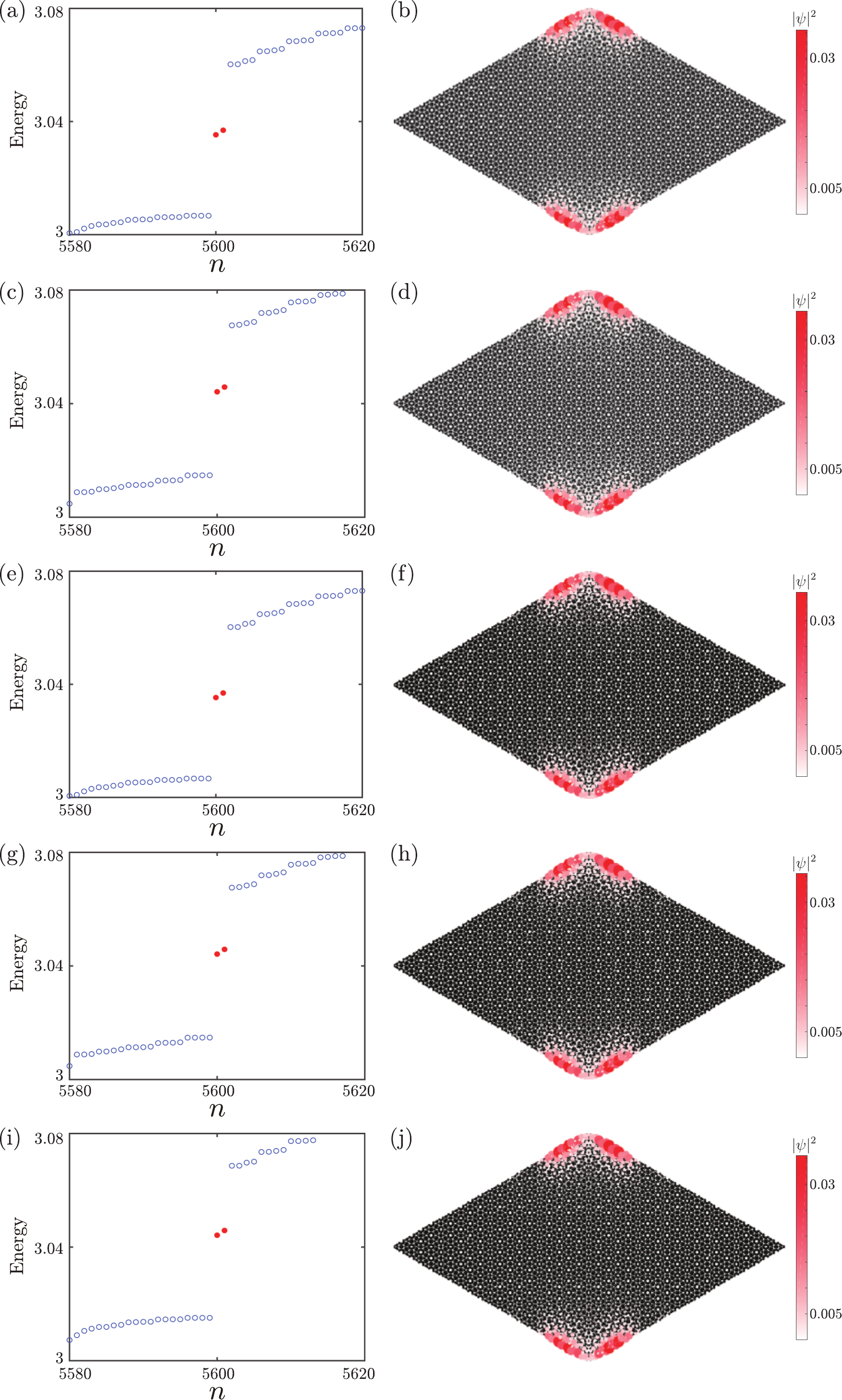} \caption{Magnon energy spectrum of the Hamiltonian~(\ref{HS3}) of the rhombus-shaped TBHM sample versus the eigenvalue index $n$, where (a) NN and NNN intralayer FM exchange coupling and NN interlayer FM exchange coupling, (c) NN and NNN intralayer FM exchange coupling and NN and NNN interlayer FM exchange coupling, (e) NN, NNN and NNNN intralayer FM exchange coupling and NN interlayer FM exchange coupling, (g) NN, NNN and NNNN intralayer FM exchange coupling and NN and NNN interlayer FM exchange coupling, (i) NN, NNN and NNNN intralayer FM exchange coupling and NN, NNN and NNNN interlayer FM exchange coupling. (b), (d), (f), (h), and (j) The spatial distribution of the probability density of the two in-gap states in (a), (c), (e), (g), and (i), respectively. The color map shows the values of the probability density. We take the exchange coupling ratio $J_{\bot }/J=0.2$, $\delta=0.028$, $a_{0}=1$, $b_{0}=1$ and the lattice site number $N=11200$.}%
	\label{figS2}
\end{figure}

Next, we transform the long-range spin Hamiltonian to the magnon Hamiltonian by using the Holstein-Primakoff (HP) transformation, and the effective magnon Hamiltonian is expressed as
\begin{eqnarray}\label{HS3}
H &=&\left( 3J_{1}S+6J_{2}S+3J_{3}S\right) \sum_{i,l}d_{i,l}^{\dag
}d_{i,l}-J_{1}S\sum_{\left\langle i,j\right\rangle ,l}\left( d_{i,l}^{\dag
}d_{j,l}+d_{j,l}^{\dag }d_{i,l}\right) \nonumber  \\
&&-J_{2}S\sum_{\left\langle \left\langle i,j\right\rangle \right\rangle
,l}\left( d_{i,l}^{\dag }d_{j,l}+d_{j,l}^{\dag }d_{i,l}\right)
-J_{3}S\sum_{\left\langle \left\langle \left\langle i,j\right\rangle
\right\rangle \right\rangle ,l}\left( d_{i,l}^{\dag }d_{j,l}+d_{j,l}^{\dag
}d_{i,l}\right) \nonumber  \\
&&-\sum_{\left\langle i,j\right\rangle }J_{\perp }^{1}S\left[ \left(
d_{i,2}^{\dag }d_{j,1}+d_{j,1}^{\dag }d_{i,2}\right) -\left( d_{i,2}^{\dag
}d_{i,2}+d_{j,1}^{\dag }d_{j,1}\right) \right]  \\
&&-\sum_{\left\langle \left\langle i,j\right\rangle \right\rangle }J_{\perp
}^{2}S\left[ \left( d_{i,2}^{\dag }d_{j,1}+d_{j,1}^{\dag }d_{i,2}\right)
-\left( d_{i,2}^{\dag }d_{i,2}+d_{j,1}^{\dag }d_{j,1}\right) \right]  \nonumber  \\
&&-\sum_{\left\langle \left\langle \left\langle i,j\right\rangle
\right\rangle \right\rangle }J_{\perp }^{3}S\left[ \left( d_{i,2}^{\dag
}d_{j,1}+d_{j,1}^{\dag }d_{i,2}\right) -\left( d_{i,2}^{\dag
}d_{i,2}+d_{j,1}^{\dag }d_{j,1}\right) \right]  \nonumber
\end{eqnarray}
where $J_{1}=J_{0}$, $J_{2}=J_{0}\exp \left( -\frac{\sqrt{3}-a_{0}}{\delta }\right)$, $J_{3}=J_{0}\exp \left( -\frac{2-a_{0}}{\delta }\right)$, $J_{\perp }^{1}=J_{\perp }^{0}$, $J_{\perp }^{2}=J_{\perp }^{0}\exp \left(-\frac{b_{2}-b_{0}}{\delta }\right)$, and $J_{\perp}^{3}=J_{\perp }^{0}\exp \left(-\frac{b_{3}-b_{0}}{\delta }\right)$. In Fig.~\ref{figS2}, we show the magnon corner states obtained by the long-range effective magnon Hamiltonian Eq.~(\ref{HS3}) for a rhombus-shaped sample with different long-range the intralayer and interlayer coupling terms, respectively. Here, the magnon corner states in the TBHMs can be found in the following three cases: the Hamiltonian with remote intralayer and NN interlayer FM exchange couplings, the Hamiltonian with NN intralayer and remote interlayer FM exchange couplings, and the Hamiltonian with remote intralayer and remote interlayer FM exchange couplings. Furthermore, we can see that the magnon corner states still exist but experience an energy shift.

\section{$k$-space spin wave Hamiltonian}

At the commensurate angle, the original translational symmetry of AA-stacked bilayer honeycomb lattice is broken, but the moir{\'e} translational symmetry can be defined and used to display the periodicity of the system. The two unit vectors of the superlattice are $\mathbf{L}_{1}\!=\!m\mathbf{a}_{1}+n\mathbf{a}_{2}$ and $\mathbf{L}_{2}\!=\!-n\mathbf{a}_{1}+( m+n) \mathbf{a}_{2}$, where $\mathbf{a}_{1}\!=\!(\frac{3\sqrt{7}}{7},-\frac{2\sqrt{21}}{7})$, $\mathbf{a}_{2}\!=\!(\frac{9\sqrt{7}}{14},\frac{\sqrt{21}}{14})$. The formation of the moir{\'e} superlattice depends on the twist angle. For any coprime integers $m$ and $n$, a twist angle can be written as
\begin{equation}
\theta =2\arcsin \frac{m-n}{2\sqrt{m^{2}+n^{2}+mn}},
\end{equation}
and a unit cell contains $N=4(m^2+mn+n^2)$ sites. The minimum site number is $28$, when $m=2$, $n=1$, and the correspoinding twist angle is $\theta=21.78^{\circ}$. The basis vectors of the supercell are $\mathbf{L}_{1} =( \frac{3}{2}\sqrt{7},-\frac{1}{2}\sqrt{21})$, $\mathbf{L}_{2} =( \frac{3}{2}\sqrt{7},\frac{1}{2}\sqrt{21})$.

Thereby, under the Fourier transformation, the magnon Hamiltonian Eq.~(\ref{HS3}) transfers into a $28 \times 28$ magnon Hamiltonian in the $k$-space, which reads
\begin{equation}
H=\sum_{\mathbf{k}} \Psi _{\mathbf{k}}^{\dag } H_{\mathbf{k}} \Psi _{\mathbf{k}},
\end{equation}
with the basis $\Psi _{\mathbf{k}}^{\dag }=( c_{\mathbf{k},1}^{\dag },\cdots ,c_{\mathbf{k},28}^{\dag })$. The numbers in the basis correspond to the site marks in the unit cell, as shown in Fig.~\ref{figS1}. Here, we ignore the long-range interlayer couplings for simplicity. Then, $H_{\mathbf{k}}$ can be expressed as
\begin{equation}\label{SHKFM}
H_{\mathbf{k}}=\left(
\begin{array}{cc}
H_{11} & H_{12} \\
H_{21} & H_{22}%
\end{array}\right) S +( 3J_{1}+6J_{2}+3J_{3})S \mathcal{I}_{28},
\end{equation}
with $H_{11}=\left( \begin{array}{cc}
H_{11}^{11} & H_{11}^{12} \\
H_{11}^{21} & H_{11}^{22}%
\end{array}\right)$, $H_{12}=\left(
\begin{array}{cc}
H_{12}^{11} & 0 \\
0 & H_{12}^{22}%
\end{array}\right) $, $H_{21}=H_{12}^{\dag }(\mathbf{k})$, and $H_{22}=\left(
\begin{array}{cc}
H_{22}^{11} & H_{22}^{12} \\
H_{22}^{21} & H_{22}^{22}
\end{array}\right)$,
where $\mathcal{I}_{28}$ is the $28 \times 28$ identity matrix. The elements of the matrices ($H_{11}$, $H_{12}$, $H_{21}$, and $H_{22}$) are all $7\times7$ matrix. The matrix $H_{11}$ in the $k$-space $28 \times 28$ magnon Hamiltonian shows the intralayer coupling of the first layer. The matrix $H_{22}$ shows the intralayer coupling of the second layer. The matrices $H_{12}$ and $H_{21}$ show the interlayer coupling between the layer $1$ and the layer $2$.

The concrete matrix forms of the elements of $H_{11}$ are
\begin{equation}
H_{11}^{11}=\left(
\begin{array}{ccccccc}
0 & -J_{1} & -J_{2} & -J_{3} & -J_{2}e^{-i\mathbf{k}\cdot\mathbf{L_{1}}} & -J_{1}e^{-i\mathbf{k}\cdot\mathbf{L_{1}}} & -J_{2}e^{-i\mathbf{k}\cdot\mathbf{L_{1}}} \\
-J_{1} & 0 & -J_{1} & -J_{2} & 0 & -J_{2}e^{-i\mathbf{k}\cdot\mathbf{L_{1}}} & -J_{3}e^{-i\mathbf{k}\cdot\mathbf{L_{1}}} \\
-J_{2} & -J_{1} & J_{\perp} & -J_{1} & -J_{2} & -J_{3} & -J_{2} \\
-J_{3} & -J_{2} & -J_{1} & 0 & -J_{1} & -J_{2} & -J_{3} \\
-J_{2}e^{i\mathbf{k}\cdot\mathbf{L_{1}}} & 0 & -J_{2} & -J_{1} & 0 & -J_{1} & -J_{2} \\
-J_{1}e^{i\mathbf{k}\cdot\mathbf{L_{1}}} & -J_{2}e^{i\mathbf{k}\cdot\mathbf{L_{1}}} & -J_{3} & -J_{2} & -J_{1} & 0 & -J_{1} \\
-J_{2}e^{i\mathbf{k}\cdot\mathbf{L_{1}}} & -J_{3}e^{i\mathbf{k}\cdot\mathbf{L_{1}}} & -J_{2} & -J_{3} & -J_{2} & -J_{1} & 0%
\end{array}%
\right),
\end{equation}
\begin{equation}
H_{11}^{12}=\left(
\begin{array}{ccccccc}
0 & -J_{2}e^{-i\mathbf{k}\cdot\mathbf{L_{2}}} & -J_{1}e^{-i\mathbf{k}\cdot\mathbf{L_{2}}} & -J_{2}e^{-i\mathbf{k}\cdot\mathbf{L_{2}}} & -J_{3}e^{-i\mathbf{k}\cdot\mathbf{L_{1}}} & -J_{2}e^{-i\mathbf{k}\cdot\mathbf{L_{1}}} & -J_{3}e^{-i\mathbf{k}\cdot(\mathbf{L_{1}+L_{2}})} \\
-J_{2} & -J_{3} & -J_{2}e^{-i\mathbf{k}\cdot\mathbf{L_{2}}} & -J_{3}e^{-i\mathbf{k}\cdot\mathbf{L_{2}}} & -J_{2}e^{-i\mathbf{k}\cdot\mathbf{L_{1}}} & -J_{1}e^{-i\mathbf{k}\cdot\mathbf{L_{1}}} & -J_{2}e^{-i\mathbf{k}\cdot\mathbf{L_{1}}} \\
-J_{1} & -J_{2} & -J_{3}e^{-i\mathbf{k}\cdot\mathbf{L_{2}}} & -J_{2}e^{-i\mathbf{k}\cdot\mathbf{L_{2}}} & 0 & -J_{2}e^{-i\mathbf{k}\cdot\mathbf{L_{1}}} & -J_{3}e^{-i\mathbf{k}\cdot\mathbf{L_{1}}} \\
-J_{2} & 0 & -J_{2}e^{-i\mathbf{k}\cdot\mathbf{L_{2}}} & -J_{1}e^{-i\mathbf{k}\cdot\mathbf{L_{2}}} & -J_{2}e^{-i\mathbf{k}\cdot\mathbf{L_{2}}} & -J_{3}e^{-i\mathbf{k}\cdot\mathbf{L_{2}}} & -J_{2}e^{-i\mathbf{k}\cdot\mathbf{L_{2}}} \\
-J_{3} & -J_{2}e^{-i\mathbf{k}\cdot(\mathbf{L_{2}-L_{1}})} & -J_{3}e^{-i\mathbf{k}\cdot(\mathbf{L_{2}-L_{1}})} & -J_{2}e^{-i\mathbf{k}\cdot\mathbf{L_{2}}} & -J_{3}e^{-i\mathbf{k}\cdot\mathbf{L_{2}}} & -J_{2}e^{-i\mathbf{k}\cdot\mathbf{L_{2}}} & -J_{1}e^{-i\mathbf{k}\cdot\mathbf{L_{2}}} \\
-J_{2} & -J_{3}e^{-i\mathbf{k}\cdot(\mathbf{L_{2}-L_{1}})} & -J_{2}e^{-i\mathbf{k}\cdot(\mathbf{L_{2}-L_{1}})} & 0 & -J_{2} & -J_{3} & -J_{2}e^{-i\mathbf{k}\cdot\mathbf{L_{2}}} \\
-J_{1} & -J_{2} & -J_{3} & -J_{2} & -J_{1} & -J_{2} & 0%
\end{array}%
\right),
\end{equation}

\begin{equation}
H_{11}^{21}=(H_{11}^{12})^{\dag },
\end{equation}
and
\begin{equation}
H_{11}^{22}=\left(
\begin{array}{ccccccc}
 0 & -J_{1} & -J_{2} & -J_{3} & -J_{2} & -J_{3}e^{-i\mathbf{k}\cdot\mathbf{L_{1}}} & -J_{2}e^{-i\mathbf{k}\cdot\mathbf{L_{1}}} \\
 -J_{1} & 0 & -J_{1} & -J_{2} & -J_{3} & -J_{2}e^{-i\mathbf{k}\cdot\mathbf{L_{1}}} & -J_{1}e^{-i\mathbf{k}\cdot\mathbf{L_{1}}} \\
 -J_{2} & -J_{1} & 0 & -J_{1} & -J_{2} & 0 & -J_{2}e^{-i\mathbf{k}\cdot\mathbf{L_{1}}} \\
 -J_{3} & -J_{2} & -J_{1} & 0 & -J_{1} & -J_{2} & -J_{3} \\
 -J_{2} & -J_{3} & -J_{2} & -J_{1} & J_{\perp} & -J_{1} & -J_{2} \\
 -J_{3}e^{i\mathbf{k}\cdot\mathbf{L_{1}}} & -J_{2}e^{i\mathbf{k}\cdot\mathbf{L_{1}}} & 0 & -J_{2} &-J_{1} & 0 & -J_{1} \\
 -J_{2}e^{i\mathbf{k}\cdot\mathbf{L_{1}}} & -J_{1}e^{i\mathbf{k}\cdot\mathbf{L_{1}}} & -J_{2}e^{i\mathbf{k}\cdot\mathbf{L_{1}}} & -J_{3} & -J_{2} & -J_{1} & 0%
\end{array}%
\right).
\end{equation}
The concrete matrix forms of the elements of $H_{12}$ are
\begin{equation}
H_{12}^{11}=\left(
\begin{array}{ccccccc}
0 & 0 & 0 & 0 & 0 & 0 & 0 \\
0 & 0 & 0 & 0 & 0 & 0 & 0 \\
0 & 0 & -J_{\perp} & 0 & 0 & 0 & 0 \\
0 & 0 & 0 & 0 & 0 & 0 & 0 \\
0 & 0 & 0 & 0 & 0 & 0 & 0 \\
0 & 0 & 0 & 0 & 0 & 0 & 0 \\
0 & 0 & 0 & 0 & 0 & 0 & 0%
\end{array}%
\right),
\end{equation}
and
\begin{equation}
H_{12}^{22}=\left(
\begin{array}{ccccccc}
0 & 0 & 0 & 0 & 0 & 0 & 0 \\
0 & 0 & 0 & 0 & 0 & 0 & 0 \\
0 & 0 & 0 & 0 & 0 & 0 & 0 \\
0 & 0 & 0 & 0 & 0 & 0 & 0 \\
0 & 0 & 0 & 0 & -J_{\perp} & 0 & 0 \\
0 & 0 & 0 & 0 & 0 & 0 & 0 \\
0 & 0 & 0 & 0 & 0 & 0 & 0%
\end{array}%
\right),
\end{equation}
where $J_{\perp}$ in the diagonal part comes from the $z$-component of the spin-spin interaction term. The concrete matrix forms of the elements of $H_{22}$ are
\begin{equation}
H_{22}^{11}=\left(
\begin{array}{ccccccc}
0 & -J_{1} & -J_{2} & -J_{3} & -J_{2}e^{-i\mathbf{k}\cdot\mathbf{L_{2}}} & -J_{1}e^{-i\mathbf{k}\cdot\mathbf{L_{2}}} & -J_{2}e^{-i\mathbf{k}\cdot\mathbf{L_{2}}} \\
-J_{1} & 0 & -J_{1} & -J_{2} & 0 & -J_{2}e^{-i\mathbf{k}\cdot\mathbf{L_{2}}} & -J_{3}e^{-i\mathbf{k}\cdot\mathbf{L_{2}}} \\
-J_{2} & -J_{1} & J_{\perp} & -J_{1} & -J_{2} & -J_{3} & -J_{2} \\
-J_{3} & -J_{2} & -J_{1} & 0 & -J_{1} & -J_{2} & -J_{3} \\
-J_{2}e^{i\mathbf{k}\cdot\mathbf{L_{2}}} & 0 & -J_{2} & -J_{1} & 0 & -J_{1} & -J_{2} \\
-J_{1}e^{i\mathbf{k}\cdot\mathbf{L_{2}}} & -J_{2}e^{i\mathbf{k}\cdot\mathbf{L_{2}}} & -J_{3} & -J_{2} & -J_{1} & 0 & -J_{1} \\
-J_{2}e^{i\mathbf{k}\cdot\mathbf{L_{2}}} & -J_{3}e^{i\mathbf{k}\cdot\mathbf{L_{2}}} & -J_{2} & -J_{3} & -J_{2} & -J_{1} & 0%
\end{array}%
\right),
\end{equation}
\begin{equation}
H_{22}^{12}=\left(
\begin{array}{ccccccc}
0 & -J_{2}e^{-i\mathbf{k}\cdot\mathbf{L_{1}}} & -J_{1}e^{-i\mathbf{k}\cdot\mathbf{L_{1}}} & -J_{2}e^{-i\mathbf{k}\cdot\mathbf{L_{1}}} & -J_{3}e^{-i\mathbf{k}\cdot\mathbf{L_{2}}} & -J_{2}e^{-i\mathbf{k}\cdot\mathbf{L_{2}}} & -J_{3}e^{-i\mathbf{k}\cdot(\mathbf{L_{1}+L_{2}})} \\
-J_{2} & -J_{3} & -J_{2}e^{-i\mathbf{k}\cdot\mathbf{L_{1}}} & -J_{3}e^{-i\mathbf{k}\cdot\mathbf{L_{1}}} & -J_{2}e^{-i\mathbf{k}\cdot\mathbf{L_{2}}} & -J_{1}e^{-i\mathbf{k}\cdot\mathbf{L_{2}}} & -J_{2}e^{-i\mathbf{k}\cdot\mathbf{L_{2}}} \\
-J_{1} & -J_{2} & -J_{3}e^{-i\mathbf{k}\cdot\mathbf{L_{1}}} & -J_{2}e^{-i\mathbf{k}\cdot\mathbf{L_{1}}} & 0 & -J_{2}e^{-i\mathbf{k}\cdot\mathbf{L_{2}}} & -J_{3}e^{-i\mathbf{k}\cdot\mathbf{L_{2}}} \\
-J_{2} & 0 & -J_{2}e^{-i\mathbf{k}\cdot\mathbf{L_{1}}} & -J_{1}e^{-i\mathbf{k}\cdot\mathbf{L_{1}}} & -J_{2}e^{-i\mathbf{k}\cdot\mathbf{L_{1}}} & -J_{3}e^{-i\mathbf{k}\cdot\mathbf{L_{1}}} & -J_{2}e^{-i\mathbf{k}\cdot\mathbf{L_{1}}} \\
-J_{3} & -J_{2}e^{i\mathbf{k}\cdot(\mathbf{L_{2}-L_{1}})} & -J_{3}e^{i\mathbf{k}\cdot(\mathbf{L_{2}-L_{1}})} & -J_{2}e^{-i\mathbf{k}\cdot\mathbf{L_{1}}} & -J_{3}e^{-i\mathbf{k}\cdot\mathbf{L_{1}}} & -J_{2}e^{-i\mathbf{k}\cdot\mathbf{L_{1}}} & -J_{1}e^{-i\mathbf{k}\cdot\mathbf{L_{1}}} \\
-J_{2} & -J_{3}e^{i\mathbf{k}\cdot(\mathbf{L_{2}-L_{1}})} & -J_{2}e^{i\mathbf{k}\cdot(\mathbf{L_{2}-L_{1}})} & 0 & -J_{2} & -J_{3} & -J_{2}e^{-i\mathbf{k}\cdot\mathbf{L_{1}}} \\
-J_{1} & -J_{2} & -J_{3} & -J_{2} & -J_{1} & -J_{2} & 0%
\end{array}%
\right),
\end{equation}

\begin{equation}
H_{22}^{21}=(H_{22}^{12})^{\dag },
\end{equation}
and
\begin{equation}
H_{22}^{22}=\left(
\begin{array}{ccccccc}
 0 & -J_{1} & -J_{2} & -J_{3} & -J_{2} & -J_{3}e^{-i\mathbf{k}\cdot\mathbf{L_{2}}} & -J_{2}e^{-i\mathbf{k}\cdot\mathbf{L_{2}}} \\
 -J_{1} & 0 & -J_{1} & -J_{2} & -J_{3} & -J_{2}e^{-i\mathbf{k}\cdot\mathbf{L_{2}}} & -J_{1}e^{-i\mathbf{k}\cdot\mathbf{L_{2}}} \\
 -J_{2} & -J_{1} & 0 & -J_{1} & -J_{2} & 0 & -J_{2}e^{-i\mathbf{k}\cdot\mathbf{L_{2}}} \\
 -J_{3} & -J_{2} & -J_{1} & 0 & -J_{1} & -J_{2} & -J_{3} \\
 -J_{2} & -J_{3} & -J_{2} & -J_{1} & J_{\perp} & -J_{1} & -J_{2} \\
 -J_{3}e^{i\mathbf{k}\cdot\mathbf{L_{2}}} & -J_{2}e^{i\mathbf{k}\cdot\mathbf{L_{2}}} & 0 & -J_{2} &-J_{1} & 0 & -J_{1} \\
 -J_{2}e^{i\mathbf{k}\cdot\mathbf{L_{2}}} & -J_{1}e^{i\mathbf{k}\cdot\mathbf{L_{2}}} & -J_{2}e^{i\mathbf{k}\cdot\mathbf{L_{2}}} & -J_{3} & -J_{2} & -J_{1} & 0%
\end{array}%
\right).
\end{equation}

\begin{figure}[t]
	\includegraphics[width=16cm]{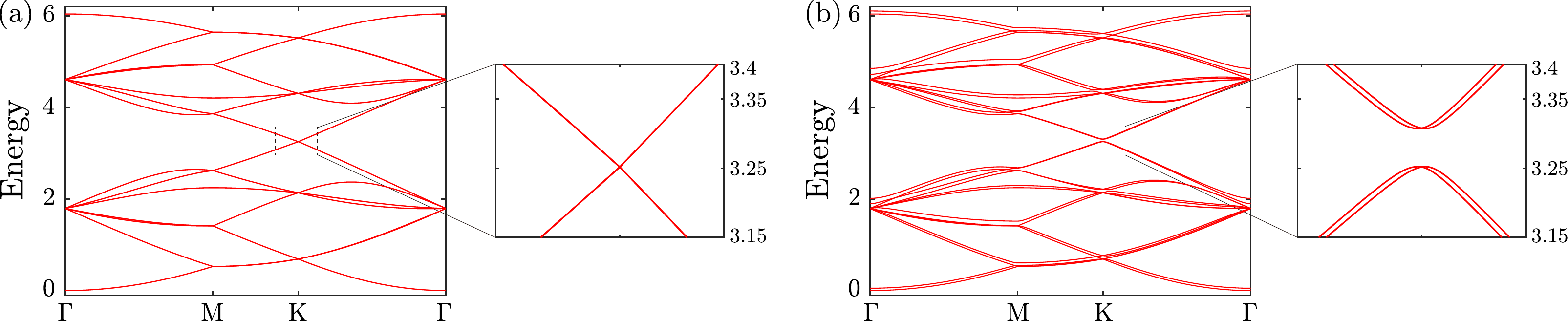} \caption{FM magnon band structures of the long-range TBHMs with $\theta=21.78^{\circ}$ for (a) $J_{\bot }/J_{0}=0$ and (b) $J_{\bot }/J_{0}=0.2$. The zoom-ins show the dispersion around the $K$ point. We take the model parameters $\delta=0.5$ and $a_{0}=1$.}%
\label{figS3}
\end{figure}

The magnon band structure of the bulk state of TBHM obtained by diagonalizing the long-range magnon Hamiltonian Eq.~(\ref{SHKFM}) is shown in Fig.~\ref{figS3}. Fig.~\ref{figS3}(a) shows the magnon band structure in the absence of interlayer interaction. We present the magnon band structures of the  TBHM system with a finite interlayer interaction in Fig.~\ref{figS3}(b), where we set $J_{\bot }/J_{0}=0.2$. We can see that the magnon band structure possesses a narrow energy gap induced when turning on the interlayer interaction. The magnon corner states residing in the topological gap has been displayed in Fig.~\ref{figS2}.

\section{TBHM coupled by AFM exchange interaction}
We briefly discuss the case of antiferromagnetic (AFM) interlayer Heisenberg interaction in this section. We assume that the spins of the first (second) layer are along the positive (negative) $z$-direction. The HP transformation in the first layer is
\begin{align}
S_{i,1}^{+}=&S^{x}_{i,1}+iS^{y}_{i,1}\simeq \sqrt{2S}d_{i,1}, \nonumber \\
 S_{i,1}^{-}=&S^{x}_{i,1}-iS^{y}_{i,1}\simeq \sqrt{2S} d_{i,1}^{\dag }, \\
 S_{i,1}^{z}=&S-d_{i,1}^{\dag }d_{i,1}, \nonumber
\end{align}
and HP transformation in the second layer is
\begin{align}
S_{i,2}^{+}=&S^{x}_{i,2}+iS^{y}_{i,2}\simeq \sqrt{2S}d_{i,2}^{\dag }, \nonumber \\
 S_{i,2}^{-}=&S^{x}_{i,2}-iS^{y}_{i,2}\simeq \sqrt{2S} d_{i,2}, \\
 S_{i,2}^{z}=&d_{i,2}^{\dag }d_{i,2}-S. \nonumber
\end{align}
Then, the effective magnon Hamiltonian reads
\begin{align}
H_{\rm AFM}=&\left ( 3J_{1}S+6J_{2}S+3J_{3}S \right )\sum_{i}(d_{i,\rm 1}^{\dag}d_{i,\rm 1}+d_{i,\rm 2}d_{i,\rm 2}^{\dag})-J_{\perp}S\sum_{\left \langle i,j \right \rangle}\left ( d_{i,\rm 2}^{\dag}d_{i,\rm 2}+d_{j,\rm 1}d_{j,\rm 1}^{\dag} \right )\nonumber \\
&-J_{1}S\sum_{\left \langle i,j \right \rangle,l}\left( d_{i,l}d_{j,l}^{\dag}+d_{i,l}^{\dag}d_{j,l} \right)-J_{2}S\sum_{\left \langle \left \langle i,j \right \rangle \right \rangle,l}\left( d_{i,l}d_{j,l}^{\dag}+d_{i,l}^{\dag}d_{j,l} \right) \\
&-J_{3}S\sum_{\left \langle \left \langle \left \langle i,j \right \rangle \right \rangle \right \rangle,l}\left( d_{i,l}d_{j,l}^{\dag}+d_{i,l}^{\dag}d_{j,l} \right)-J_{\perp}S\sum_{\left \langle i,j \right \rangle}\left( d_{i,\rm 2}^{\dag }d_{j,\rm 1}^{\dag}+d_{i,\rm 2}d_{j,\rm 1} \right) \nonumber.
\end{align}

\begin{figure}[t]
	\includegraphics[width=16cm]{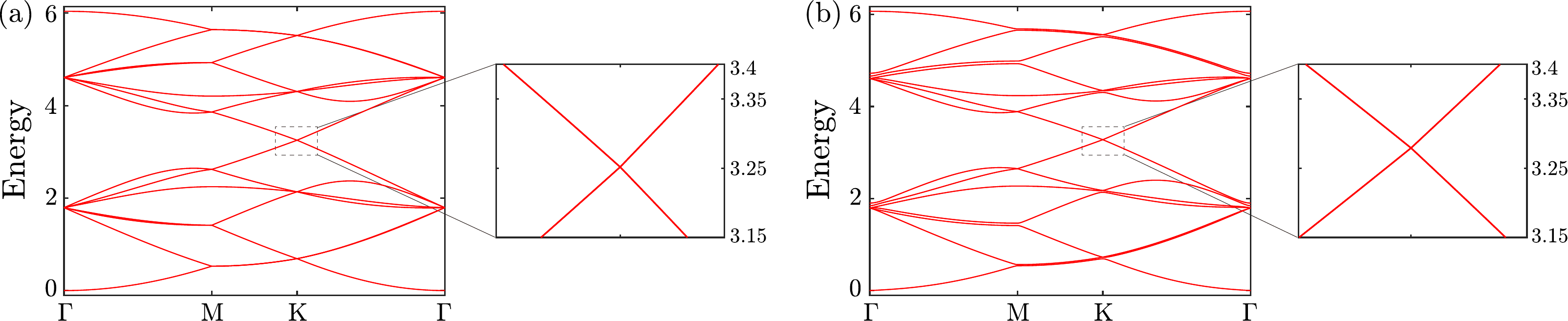} \caption{AFM magnon band structures of the long-range TBHMs with $\theta=21.78^{\circ}$ for (a) $J_{\bot }/J_{0}=0$ and (b) $J_{\bot }/J_{0}=-0.2$. A zoom of the region near K point indicated by a dashed black box in the left panel is shown in the right panel. We take the model parameters $\delta=0.5$ and $a_{0}=1$.}%
	\label{figS4}
\end{figure}

Again, in terms of the Fourier transformation, we obtain a $28 \times 28$ $k$-space magnon Hamiltonian, which can be expressed as $H_{\rm AFM}=\sum_{\mathbf{k}} \psi _{\mathbf{k}}^{\dag } H_{\rm AFM}(\mathbf{k}) \psi _{\mathbf{k}}$ with the basis $\psi_{\mathbf{k}}=(c_{\mathbf{k},1},...,c_{\mathbf{k},14},c_{\mathbf{-k},15}^{\dag},...,c_{\mathbf{-k},28}^{\dag})^{T}$. The difference between $H_{\rm AFM}(\mathbf{k})$ and $H_{\mathbf{k}}$ is that $J_{\bot }$ in $H_{11}$ and $H_{22}$ of $H_{\mathbf{k}}$ is replaced by $-J_{\bot }$. To reveal the AFM magnon band structures, we use the paraunitary Bogoliubov transformation $\psi_{\mathbf{k}}=R(\mathbf{k})\phi_{\mathbf{k}}$ to diagonalize the $k$-space magnon Hamiltonian as $R(\mathbf{k})^{\dag}H_{\rm AFM}(\mathbf{k})R(\mathbf{k})\!=\!D$, where $D$ is a diagonal matrix and $\phi_{\mathbf{k}}=(1,...,1,-1,...,-1)^{T}$. Figure~\ref{figS4}(a) shows the magnon band structure of the long-range TBHM system without interlayer interaction. Subsequently, we present the magnon band structures of the long-range TBHM system with a finite AFM interlayer interaction in Fig.~\ref{figS4}(b), where we set $J_{\bot }/J_{0}=-0.2$. We find that the AFM magnon energy band keeps gapless regardless of whether the AFM interlayer interaction exists, which is in contrast to the FM interlayer coupling case.

\section{Six-fold rotation symmetric magnon corner states}

\begin{figure}[t]
	\includegraphics[width=10cm]{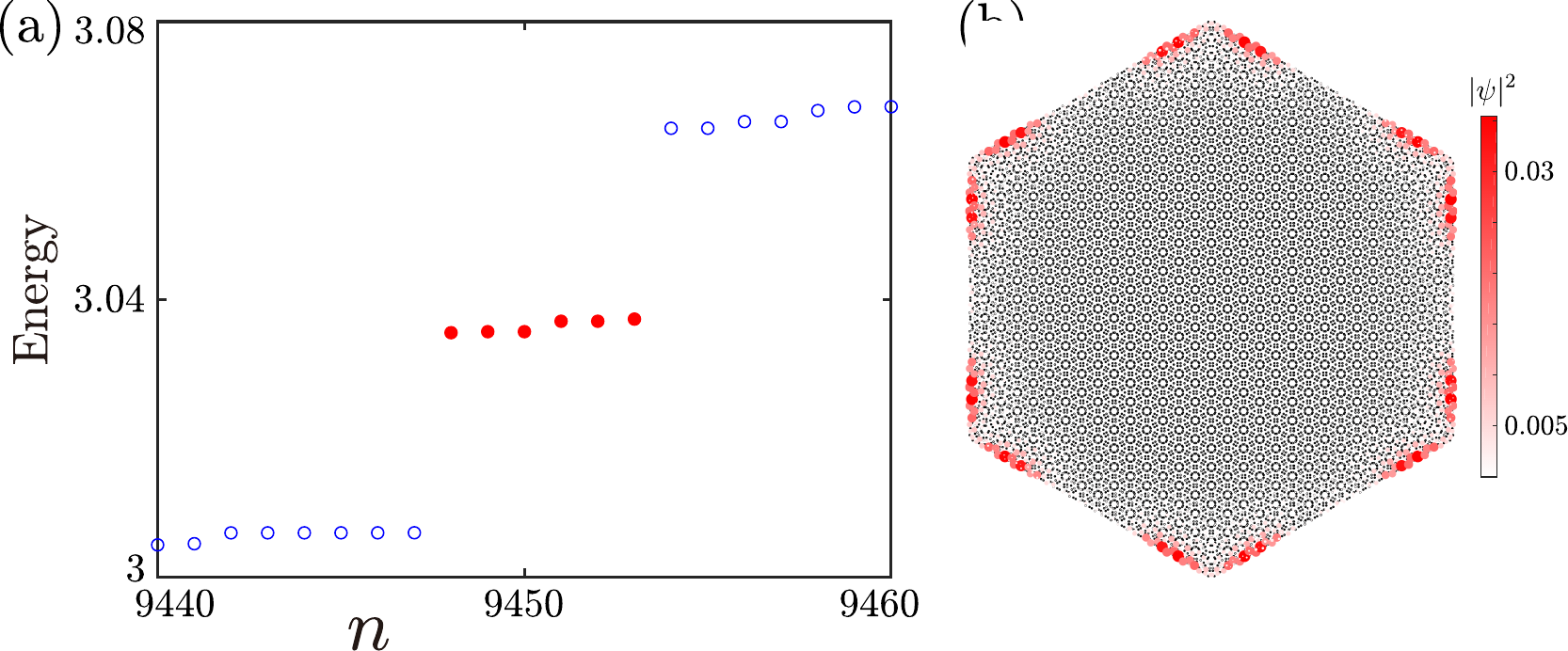} \caption{(a) Magnon energy spectrum of the Hamiltonian Eq.~(3) in the main text of the TBHM system with regular hexagonal boundary versus the eigenvalue index $n$. Red dots mark all the in-gap states. (b) The probability density of the six in-gap states in (a). The color map shows the values of the probability density. We take the model parameters $J/J=1$, $J_{\bot }/J=0.2$, and the lattice site number $N=18900$.}%
	\label{figS5}
\end{figure}

In this section, we show the magnon corner states in the TBHM system with a regular hexagonal boundary shape. By numerically diagonalizing the magnon Hamiltonian Eq.~(3) in the main text under a hexagonal boundary, we plot the magnon energy spectrum versus the eigenvalue index $n$ and the spatial probability density of the in-gap states in Fig.~\ref{figS5}. It is found that the magnon energy spectrum shows an energy gap induced by the interlayer FM coupling, and even more interestingly six in-gap states reside in the energy gap. We can also see that the six in-gap states are symmetrically localized at the six corners of the regular hexagon. The sixfold rotation symmetric in-gap corner states are a hallmark feature of the SOTMI in the TBHM.

\begin{figure}[t]
	\includegraphics[width=12cm]{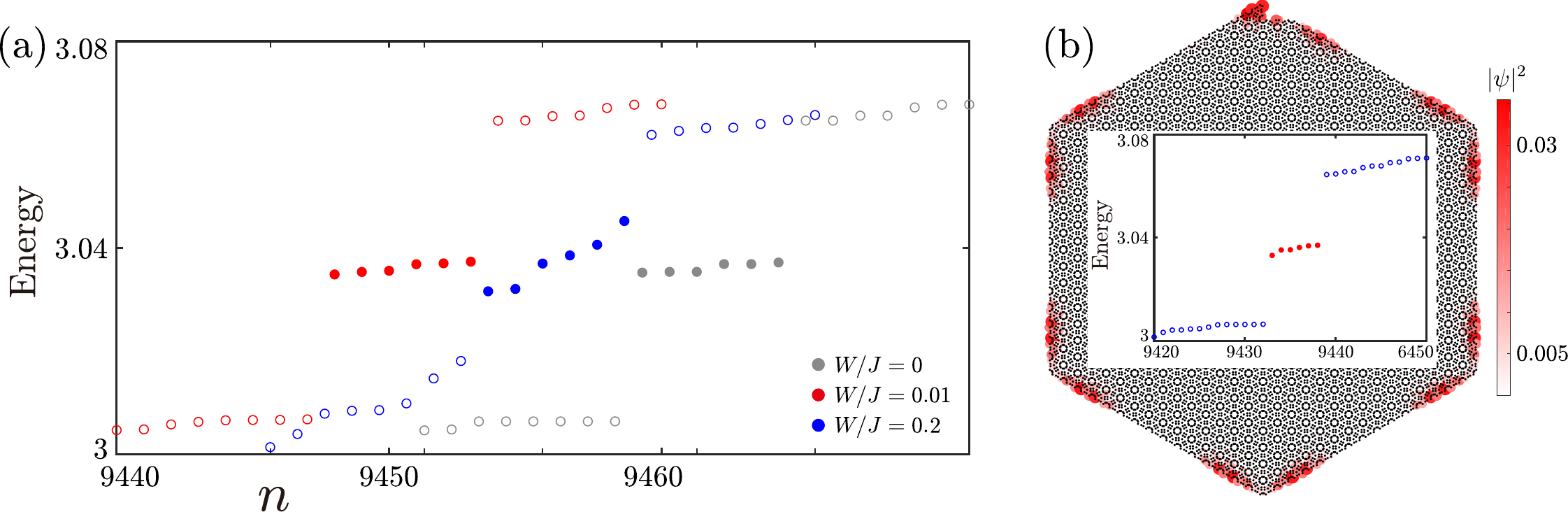} \caption{(a) Magnon energy spectrum of the total Hamiltonian $H+H_{z}$ versus the eigenvalue index $n$. Red dots mark all the in-gap states. For comparison, we also plot the magnon energy spectrum (marked by gray circles and dots) without disorder ($W/J=0$) and the magnon energy spectrum (marked by blue circles and dots) with strong disorder ($W/J=0.2$). We take the model parameters $J_{\bot }/J=0.2$, $W/J=0.01$ and lattice site number $N=18900$. (b) Magnon energy spectrum of the Hamiltonian $H$ of the TBHM system with a local defect versus the eigenvalue index $n$, and the probability density of the six in-gap states, where $J_{\bot }/J=0.2$ and the lattice site number $N=18870$. Red dots mark all the in-gap states. The color map shows the values of the probability density.}%
\label{figS6}
\end{figure}

Similarly, we use a random magnetic field to examine the robustness of the rotation symmetric magnon corner states. In Fig.~\ref{figS6}(a), we plot the magnon energy spectrum versus the eigenvalue index $n$ for different disorder strengths. We find that the six in-gap magnon corner states remain stable in the presence of weak disorder, where we take the disorder strength as $W/J=0.01$, while they are destroyed by the strong disorder with the disorder strength ($W/J=0.2$) shown in Fig.~\ref{figS6}(a). In addition, we also reveal the robustness of the magnon corner states by introducing a local defect into the regular hexagonal boundary sample at the top corner, where the defect is constructed by removing $30$ sites. As shown in Fig.~\ref{figS6}(b), the six in-gap states are not degenerate, but they are still localized around the original six corners.

\begin{figure}[htb]
	\centering
	\includegraphics[width = 1.0\columnwidth] {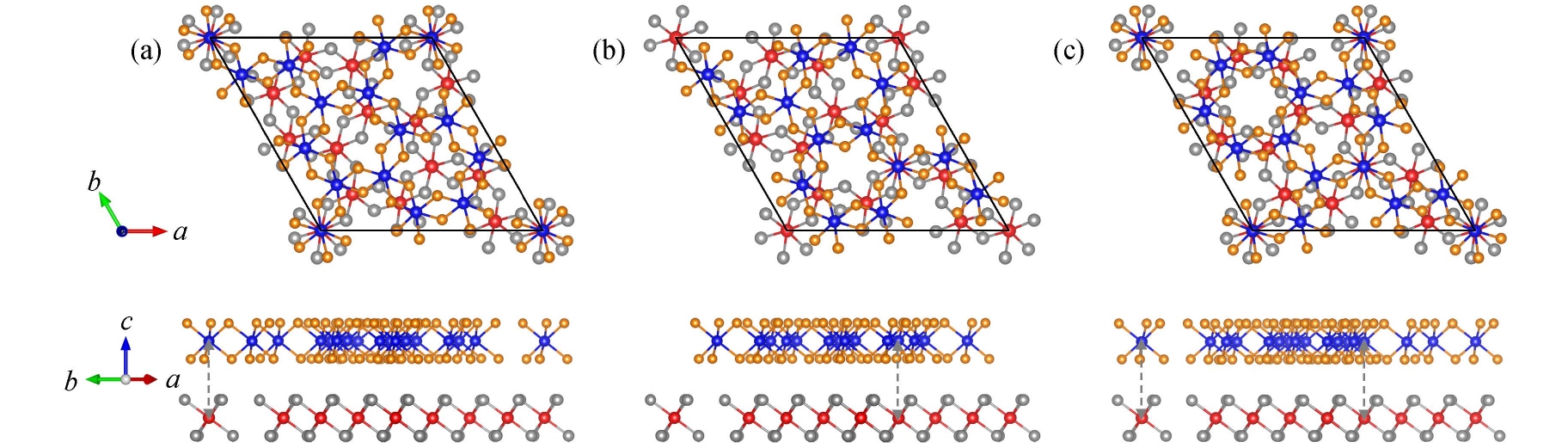}
	\caption{Top and side view of three high-symmetry configurations of twisted bilayer CrI$_3$ superlattice with twisting angle $\theta=21.7868^{\circ}$. The Blue (red) and orange (gray) balls represent Cr and I atoms of top (bottom) layer respectively.} \label{figS7}
\end{figure}

\begin{table}[htb]
	\centering
	\caption{Total energy for interlayer FM interaction (E$_\text{FM}$), AFM interaction (E$_\text{AFM}$) and their difference (E$_\text{FM}$-E$_\text{AFM}$) for three high-symmetry twisted bilayer configurations. }
	\begin{tabular*}{\hsize}{@{}@{\extracolsep{\fill}}lllllllllllll@{}}
		\toprule
		&E$_\text{FM}$(eV)&E$_\text{AFM}$(eV)&E$_\text{FM}$-E$_\text{AFM}$(meV)\\
		\colrule
		configuration (a)&-322.81887&-322.81109&-7.78\\
		configuration (b)&-322.80644&-322.80301&-3.43\\
		configuration (c)&-322.80751&-322.80029&-7.22\\
		\botrule
	\end{tabular*}
	\label{symbols}
\end{table}

\section{The first principle calculations}

The first-principles calculations were performed using the Vienna \emph{ab-initio} simulation package (VASP) \cite{Kresse1996}. The core-valence interaction was described by the frozen-core projector augmented wave (PAW) method \cite{PBE1994}. The generalized gradient approximation of Perdew-Burke-Ernzerhof (GGA-PBE) \cite{Perdew1996} was adopted for exchange-correlation functional. Van der Waals dispersion forces between the layers were accounted for through the optB88-vdW functional by using the vdW-DF method \cite{Klimes2011}. A Hubbard on-site Coulomb parameter ($U$) of 3eV was chosen for the Cr atoms to account for strong electronic correlations, as suggested by Liechtenstein et al \cite{Liechtenstein1995}. The plane-wave cutoff energy was set to 500 eV. The Brillouin zone was sampled by using a $5\times 5\times 1$ Monkhorst-Pack k-point mesh. A large vacuum layer of $20\textup{\AA }$ was used to prevent the artificial interlayer interaction. The crystal structures were fully relaxed until the force on each atom and total energy variations are smaller than $10^{-2}\textup{eV}/\textup{\AA }$ and $10^{-4}\textup{eV}/\textup{\AA }$. In the process of calculating the total energy, the energy convergence standard is set to $10^{-5}\textup{eV}/\textup{\AA }$. The obtained lattice constant of an AB stacked bilayer CrI$_3$ is $a=6.827{\AA }$.

For a twisted bilayer of $\theta=21.78^{\circ}$, there are three interlayer configurations that preserve the in-plane three-fold rotational symmetry, which are related to each other by translating $(\bm{L}_1+\bm{L}_2$)/3 and $2(\bm{L}_1+\bm{L}_2$)/3 along the long diagonal direction of the moir\'{e} superlattice, as shown in Fig.~\ref{figS7}. The interlayer magnetic interaction is related to energy difference between interlayer ferromagnetic (FM) and antiferromagnetic (AFM) spin configurations, which are listed in Tab.~\ref{symbols} for the three configurations. Interestingly, the interlayer FM energy (E$_\text{FM}$) is lower than AFM energy (E$_\text{AFM}$) for all of the three configurations, which indicates that the interlayer FM interaction is always favored. Note that, previous studies have shown that the magnon dispersion of monolayer CrI$_3$ may have a gap at the Brillouin zone corner due to anisotropic spin interactions~\cite{xu2018interplay,PhysRevB.101.060404,PhysRevLett.127.166402}, such as the Dzyaloshinskii-Moriya or Kitaev interactions, however, there is no consensus on this issue. In our work, we assume each single layer CrI$_3$ hosts gapless magnon dispersion when ignoring the interlayer couplings.

\end{document}